\documentclass{aastex}
\usepackage{spr-astr-addons}
\usepackage{url}
\usepackage[caption2]{ccaption}
\usepackage{lscape}
\RequirePackage{color}

\begin{document}

\title{Absolute Magnitude Calibration for Red Clump Stars\\}
\shorttitle{Absolute Magnitude Calibration for Red Clump Stars}
\shortauthors{S. Karaali, S. Bilir, E. Yaz G\"ok\c ce}

\author{S. Karaali \altaffilmark{1}}
\altaffiltext{1}{Istanbul University, Faculty of Science, Department 
of Astronomy and Space Sciences, 34119 University, Istanbul, Turkey\\
\email{karsa@istanbul.edu.tr}}
\and
\author{S. Bilir \altaffilmark{1}} 
\altaffiltext{1}{Istanbul University, Faculty of Science, Department 
of Astronomy and Space Sciences, 34119 University, Istanbul, Turkey\\}
\and
\author{E. Yaz G\"ok\c ce \altaffilmark{1}} 
\altaffiltext{1}{Istanbul University, Faculty of Science, Department 
of Astronomy and Space Sciences, 34119 University, Istanbul, Turkey\\}

\begin{abstract}
We combined the ($K_s$, $J-K_s$) data in \cite{Laney12} with the $V$ apparent magnitudes and trigonometric parallaxes taken from the {\em Hipparcos} catalogue and used them to fit the $M_{K_s}$ absolute magnitude to a linear polynomial in terms of $V-K_s$ colour. The mean and standard deviation of the absolute magnitude residuals,$-$0.001 and 0.195 mag, respectively, estimated for 224 red clump stars in \cite{Laney12} are (absolutely) smaller than the corresponding ones estimated by the procedure which adopts a mean $M_{K_s}=-1.613$ mag absolute magnitude for all red clump stars, $-$0.053 and 0.218 mag, respectively. The statistics estimated by applying the linear equation to the data of 282 red clump stars in \cite{Alves00} are larger, $\Delta M_{K_s}=0.209$ and $\sigma=0.524$ mag, which can be explained by a different absolute magnitude trend, i.e. condensation along a horizontal distribution.        
  
\end{abstract}

\keywords{Stars: distances, Stars: late-type, Galaxy: Solar neighbourhood} 

\section{Introduction}
Red Clump (RC) stars are core helium-burning giants. It is a prominent feature in the colour-magnitude diagrams of open clusters as well as globular clusters. Now, it is known that they are abundant in the solar neighbourhood. This sample of RC stars provides accurate absolute magnitude estimation due to their parallaxes which can be used in testing their  suitability for a distance indicator. A mean absolute magnitude in any band with small scattering can work for the purpose of the researchers. Many works has been carried out for optical and near infrared bands such as $V$, $I$ and $K_s$. Their absolute magnitudes in the optical range lie from $M_V=+0.7$ mag for those of spectral type G8 III to $M_V=+1$ mag for type K2 III \citep{Keenan99}. The absolute magnitude of these stars in the $K$ band is $M_{K_s}=-1.61\pm0.03$ mag with negligible dependence on metallicity \citep{Alves00}, but with real dispersion. \cite{Grocholski02} claimed an absolute magnitude for the RC stars rather close to the one just cited here, with some limitations however. Based on 14 open clusters, they draw the conclusion that for clusters having $-0.5<[Fe/H]\leq0$ dex and $1.58\leq t \leq 7.94$ Gyr, one can use $<M_{K_s}>=-1.61\pm0.04$ mag. 

The dependence of the $I$-band magnitude on the RC stars was extensively studied in the past from an observational point of view. In most cases the $M_I$ mean absolute magnitude is insensitive to age and metallicity \citep{Udalski98}. However, a modest variation in $M_I$ with colour and metallicity has been claimed in the literature \citep{Paczynski98, Stanek98, Sarajedini99, Zhao01, Kubiak02}. Theoretical models from \cite{Girardi01} also show a dependence of metallicity and age in $I$ band. \cite{Salaris02} stated in their study based on the models of \cite{Girardi00} that $M_K$ is a complicated function of metallicity and age. 

In a recent work, \cite{vanHelshoecht07} used the Two Micron All Sky Survey \citep[2MASS;][]{Skrutskie06} infrared data for a sample of 24 open clusters to investigate how the $K_s$-band absolute magnitude of the red clump depends on age and metallicity. They showed that a constant value of $M_{K_s}=-1.57\pm0.05$ mag is a reasonable assumption to use in distance determinations for clusters with metallicity between $-$0.5 and +0.4 dex and age between 0.31 and 7.94 Gyr. Following this work, \citet{Groenewegen08} claimed two absolute values in different bands for the RC stars, i.e. $M_{K_s}=-1.54\pm0.04$ and $M_I=-0.22\pm0.03$ mag, where the estimations were based on newly reduced the {\em Hipparcos} catalogue \cite{vanLeeuwen07}. 

In a more recent work, \citet{Laney12} determined the mean $M_K$ absolute magnitude for RC stars in the solar neighbourhood to within 2 per cent ($M_K=-1.613\pm0.015$ mag) and applied their results to the estimation of the distance of the Large Magellanic Cloud. A mean value for the $M_K$ absolute magnitude with weak or negligible dependence on metallicity makes possible to use this population as a tracer of Galactic structure and interstellar extinction, as several works have fully demonstrated in the last decade \citep[see for example][and references therein]{Lopez02, Lopez04, Cabrera05, Cabrera07a, Cabrera07b, Cabrera08, Bilir12a}.

In \citet{Bilir12b} the $M_V$, $M_J$, $M_{K_s}$ and $M_g$ absolute magnitudes of RC stars, identified by a set of constraints in the {\em Hipparcos} catalogue, were calibrated in terms of colour with $BVI$, $JHK_s$ and $gri$ photometries. In the present paper, we will focus on a single absolute magnitude, $M_{K_s}$, which is the most probable candidate for adopting as a distance indicator. Our aim is to calibrate the $M_{K_s}$ absolute magnitude as a function of colour. Thus, we expect more accurate absolute magnitudes relative to the procedure which adopts $M_{K_s}$ absolute magnitude as a constant value. The procedure is given in Section 2. The data and the relation between the colour and absolute magnitude are presented in Sections 3 and 4, respectively. The application of the procedure is devoted to Section 5, and finally a discussion is given in Section 6.

\section{The Procedure}

The absolute magnitude of a star is a function of luminosity class, temperature or colour, age and metallicity. As it is assumed that the RC stars are at the same evolutionary stage, all the RC stars are of the same luminosity class. Hence, we omit this parameter in the absolute magnitude estimation of the RC stars. Many studies are based on the {\em Hipparcos} catalogue \citep{vanLeeuwen07} which involves the solar neighbourhood stars. Using the {\em Hipparcos} catalogue is a constraint both in metallicity and age for the sample stars. This is the reason that the researchers claim weak dependence of the $M_{K_s}$ absolute magnitude on age and metallicity. The advantage of the up-dated {\em Hipparcos} catalogue \citep{vanLeeuwen07} is that the errors are smaller than the former edition. The data of open clusters are also age and metallicity constraint. For example the age and metallicity intervals for 24 open clusters in \cite{vanHelshoecht07} are $0.31\leq t\leq7.94$ Gyr, and $-0.5\leq[Fe/H]\leq+0.4$ dex, respectively. However, one can include the metal-poor globular clusters such as NGC 1261 with metallicity $[Fe/H]=-1.35$ dex into the sample used for $M_{K_s}$ absolute magnitude estimation of the RC stars. In this case, one expect metallicity- and age-dependent $M_{K_s}$ absolute magnitudes. Despite age and metallicity limitations for the data used for the $M_{K_s}$ absolute magnitude estimation, the range of the corresponding colour is large enough for considering in the estimation of the absolute magnitude in question. For example, the range of the $V-K_s$ colour of the data used in \cite{Alves00} and \cite{Laney12} is $2<V-K_s<3$ mag. Then, one should consider the $V-K_s$ colour in the $M_{K_s}$ absolute magnitude estimation. That is, we expect more accurate $M_{K_s}$ absolute magnitudes for individual RC stars rather than a constant value for the whole sample. 

\section{Data}
We used the data of \cite{Laney12} for calibration of the $M_{K_s}$ absolute magnitude in terms of $V-K_s$ colour. There are 224 RC stars in the catalogue of \cite{Laney12}. We provided the {\em Hipparcos} number, parallax, metallicity, $K_s$ magnitude, and $M_{K_s}$ absolute magnitude from the electronic version of the paper, and we added the $V$ magnitude to this catalogue to obtain the $V-K_s$ colours instead of $J-K_s$ ones. The $V$ magnitudes are taken from the {\em Hipparcos} catalogue. The relative parallax errors lie in the interval $0<\sigma_{\pi}/\pi\leq0.10$ and their median is 0.03. The parallaxes were corrected by \cite{Laney12}. The data are given in Table 1. The columns give: (1) Current number, (2) {\em Hipparcos} number, (3) the corrected parallax, (4) $[M/H]$ metallicity, (5) $V$ apparent magnitude, (6) $K_s$ apparent magnitude, (7) $V-K_s$ colour index and (8) $M_{K_s}$ absolute magnitude. As in \cite{Laney12}, we assumed no foreground reddening. Actually, the mean colour excess of 20 RC stars with {\em Hipparcos} number between 671 and 7643 in the catalogue of \citet{Laney12}, estimated by the following procedure is only $E(B-V)=0.017$ mag. The $E(B-V)$ colour excess of 20 RC stars have been evaluated in two steps. First, we used the maps of \citet{Schlegel98} and evaluated a $E_{\infty}(B-V)$ colour excess for each star. Then, we reduced them using the following procedure of \citet{Bahcall80}:
\begin{equation}
A_{d}(b)=A_{\infty}(b)\Biggl[1-\exp\Biggl(\frac{-\mid d \times \sin(b)\mid}{H}\Biggr)\Biggr].
\end{equation}
Here, $b$ and $d$ are the Galactic latitude and distance to the star, respectively. $H$ is the scale height for the interstellar dust which is adopted as 125 pc \citep{Marshall06}. $A_{\infty}(b)$ and $A_{d}(b)$ are the total absorptions for the model and for the distance to the star, respectively. $A_{\infty}(b)$ can be evaluated by means of the following equation:

\begin{equation}
A_{\infty}(b)=3.1\times E_{\infty}(B-V).
\end{equation}
$E_{\infty}(B-V)$ is the colour excess for the model taken from the \citet{Schlegel98}. Then, $E_{d}(B-V)$, i.e. the colour excess for the corresponding star at the distance $d$, can be evaluated via the equation,

\begin{equation}
E_{d}(B-V)=A_{d}(b)~/~3.1.
\end{equation}
The colour excess $E_d(B-V)$ and the classical colour excess $E(B-V)$ have the same meaning. The same case is valid for the total absorption $A_d$ and the classical absorption $A_V$. 

The metallicities are given only for 100 RC stars. The diagram for $M_{K_s}$ absolute magnitude versus $V-K_s$ colour index is given in Fig. 1. Most of the stars are concentrated in the region with $2.1<V-K_s<2.6$ and $-2<M_{K_s}<-1$ mag. However, there are about three dozen of stars beyond these limits. The extreme colours and absolute magnitudes belong to the RC stars with {\em Hipparcos} numbers 3781, 37901, 38211, 58697, 63608, 70306, 72471.

\section{Calibration of $M_{K_s}$ Absolute Magnitude to $V-K_s$ Colour}
The range of the metallicity of the sample is $-0.7<[M/H]<0.4$ dex. However, 80 per cent of them lie in a smaller metallicity interval, i.e. $-0.25\leq[M/H]\leq +0.15$ dex (Fig. 2), indicating a thin disc sample. Hence, we preferred to use the whole sample in the calibration of $M_{K_s}$ absolute magnitude to $V-K_s$ colour instead of separating it into different metallicity classes. The distribution of 80 per cent of the points in Fig. 1 is almost circular, while the complete figure gives the indication of a linear distribution. Also, the large scattering and the inhomogeneous number density hinder the selection of fitting type of the $M_{K_s}$ absolute magnitude in terms of $V-K_s$ colour. However, we considered all the points and fitted  $M_{K_s}$ to a linear equation in terms of $V-K_s$ as in the following (Fig. 3):

\begin{eqnarray}
M_{K_s}=-0.485(\pm0.065)\times (V-K_s)-0.396(\pm0.158).
\end{eqnarray}
We evaluated the $M_{K_s}$ absolute magnitude residuals, i.e. the difference between the absolute magnitude estimated by using Eq. (1) and the corresponding absolute magnitude in Table 1, and compared them with another set of absolute magnitude residuals of the same stars evaluated by adopting the value $-$1.613 mag as the $M_{K_s}$ absolute magnitude for all RC stars. This value was claimed by \citet{Alves00} and \citet{Laney12} as the mean $M_{K_s}$ absolute magnitude for RC stars. Eq. (4) is derived without considering the small foreground reddening in \citet{Laney12} stated in Section 3, i.e. $E(B-V)=0.017$ mag. If we  transform this value to the colour excess $E(V-K_s)$ by the equation $E(V-K_s)=2.74\times E(B-V)$ \citep{Karaali13}, and use it in Eq. (4) we get an absolute magnitude fainter than 0.02 mag.

The mean absolute magnitude residuals and the corresponding standard deviations for two sets are given in the third and fourth rows of Table 2. The mean of the absolute magnitude residuals evaluated by adopting the constant absolute magnitude value $M_{K_s}=-1.613$ mag, $<\Delta M_{K_s}>=$$-$0.053, is 53 times larger than the one evaluated by the linear equation, $<\Delta M_{K_s}>=$$-$0.001, in our work. Also, the standard deviation corresponding to the linear equation is smaller than the other one. However, the factor of the residuals in Fig. 4 is slightly higher than unity.
        
\section{Application of the Procedure}

We applied the procedure to the data in \cite{Alves00}. The catalogue of \cite{Alves00} involves 284 RC stars. We replaced the recent parallaxes and $K_s$ band magnitudes appeared in the {\em Hipparcos} catalogue with the old ones. The relative parallax errors lie in the interval $0<\sigma_{\pi}/\pi\leq 0.11$ and their median is 0.02. We used the following equation of \cite{Smith87} to correct the observed {\em Hipparcos} parallaxes \citep{vanLeeuwen07}:  
\begin{equation}
\pi_{0}=\pi\Biggl[\frac{1}{2}+\frac{1}{2}\sqrt{1-16(\sigma_{\pi}/\pi)^2}~\Biggr],
\end{equation}
where $\pi$ and $\pi_{0}$ are the observed and corrected parallaxes, respectively, and $\sigma_{\pi}$ denotes the error of the observed parallax. $K_{s}$ apparent magnitudes are not given for two stars, {\em Hipparcos} number: 33449, 46952. The sample is given in Table 3. The columns give (1) current number, (2) {\em Hipparcos} number, (3) $\pi_o$ corrected parallax, (4) $[M/H]$ metallicity, (5) $V$ apparent magnitude, (6) $K_s$ apparent magnitude, (7) $V-K_s$ colour index, (8) $M_{K_s}$ absolute magnitude, and (9) $Q$ parameter which indicates the quality of the $JHK_s$ magnitudes of a star. The quality of the RC stars are adopted from the {\em Hipparcos} catalogue, and the $M_{K_s}$ absolute magnitudes are evaluated by the following equation: 
\begin{equation}
M_{K_s}=K_s-10+2.1715\times\ln(\pi_0),
\end{equation}        
where $\pi_0$ is the corrected parallax of the star considered. As in \cite{Alves00}, we assumed no foreground reddening (see also Section 3). The total number of stars for which $M_{K_s}$ absolute magnitudes could be estimated is 282. 

We plotted the $M_{K_s}$ absolute magnitudes versus $V-K_s$ colours for these stars in Fig. 5. The distribution of the diagram is rather different than the one given for the data in \cite{Laney12}, i.e. there is a high condensation along a horizontal line and a large scattering beyond this formation. The range of the absolute magnitudes in Fig. 5 is much larger than the one in Fig. 1, $-3.6 \leq M_{K_s}\leq-0.3$ mag. 

We evaluated the absolute magnitudes of 282 RC stars using Eq. (4) and compared them with the original ones in Table 3. The mean of the residuals and the corresponding standard deviation are $<\Delta M_{K_s}>=0.209$ and  $\sigma=0.524$ mag. As in Section 4, we evaluated another set of statistics by adopting the value $-$1.613 mag as the $M_{K_s}$ absolute magnitude for all RC stars, i.e. $<\Delta M_{K_s}>=0.133$ and $\sigma=0.571$ mag (Table 2, row 4). Although the standard deviation evaluated by using the linear Eq. (4) is smaller than the one evaluated for the constant absolute magnitude, the corresponding mean of the residuals is about 1.6 times larger than the one estimated via constant absolute magnitude. The distribution of the residuals estimated by means of two procedures are given in Fig. 4.     

\section{Discussion}
We combined the ($K_s$, $J-K_s$) data in \cite{Laney12} with the $V$ apparent magnitudes and trigonometric parallaxes taken from the {\em Hipparcos} catalogue and used them to fit the $M_{K_s}$ absolute magnitude to a linear relation in terms of $V-K_s$ colour. We preferred the data in \cite{Laney12} for absolute magnitude calibration in terms of colour due to the high precession of the observed magnitudes. \cite{Laney12} used 0.75m telescope at South African Astronomical Observatory and observed the brightest and nearest RC stars in the solar neighbourhood which provide accurate data. 

We evaluated the $M_{K_s}$ absolute magnitude residuals, i.e. the difference between the absolute magnitude estimated by using Eq. (4) and the corresponding absolute magnitude in Table 1, and compared them with another set of absolute magnitude residuals of the same stars evaluated by adopting the value $-$1.613 mag as the $M_{K_s}$ absolute magnitude for all RC stars. This value was claimed by \citet{Alves00} and \citet{Laney12} as the mean $M_{K_s}$ absolute magnitude for RC stars. The mean of the absolute magnitude residuals evaluated by adopting the constant absolute magnitude value $M_{K_s}=-1.613$ mag, $<\Delta M_{K_s}>=$$-$0.053 mag, is 53 times larger than the one evaluated by the linear equation $<\Delta M_{K_s}>=$$-$0.001 mag in our work. Also, the standard deviation corresponding to the linear equation is smaller than the other one. This comparison shows that the $M_{K_s}$ absolute magnitudes estimated by a linear equation in terms of colour are more accurate than the constant absolute magnitudes. The result obtained from the application of the procedure to the data in \citet{Alves00} is a bit different, however. Although the standard deviation corresponding to the linear equation is smaller than the one evaluated for constant absolute magnitude, the mean of the absolute magnitude residuals estimated via linear equation is 1.6 times larger than the other one. Differences between the statistics estimated for two sets of data originate from their trends of $M_{K_s}\times (V-K_s)$ colour magnitude diagrams. The distribution of colour-magnitude diagram for the data in \citet{Laney12} is almost diagonal, whereas the absolute magnitudes of RC stars in \citet{Alves00} are condensed along a horizontal line, $M_{K_s}\sim-1.5$ mag. As one can see in the Table 3, the 2MASS data are not of best quality which probably affect the accuracy of the estimated absolute magnitudes. 

Conclusion: It has been suggested that RC stars are standard candles and a mean value of $M_{K_s}=-1.613$ mag based on the {\em Hipparcos} data has been claimed for them in the literature. In this study, we showed that the absolute magnitude for the $K_s$ band is colour dependent. However, we need data of best quality and a large sample in order to obtain a standard linear calibration of $M_{K_s}$ absolute magnitude in terms of colour.
      
\section*{Acknowledgments}
This work has been supported in part by the Scientific and Technological Research Council (T\"UB\.ITAK) 112T120. We thank to Dr. Martin Lopez-Corredoira for his comments and suggestions. This research has made use of NASA's Astrophysics Data System and the SIMBAD database, operated at CDS, Strasbourg, France.

\begin{figure}
\begin{center}
\includegraphics[scale=0.40, angle=0]{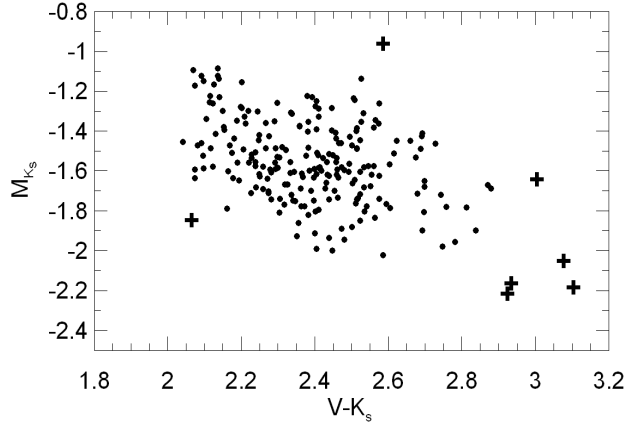}
\caption{$M_{K_s}\times(V-K_s)$ colour-absolute magnitude diagram for RC stars in Table 1. The symbol ($+$) indicates stars with large scattering.}
\label{histogram}
\end{center}
\end{figure}  

\begin{figure}
\begin{center}
\includegraphics[scale=0.40, angle=0]{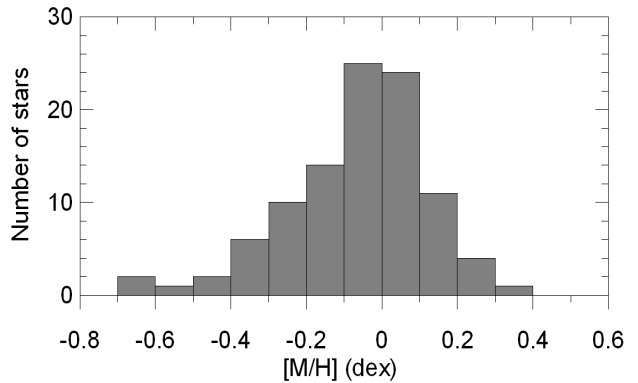}
\caption{Metallicity distribution for 100 RC stars in \cite{Laney12}.}
\label{histogram}
\end{center}
\end{figure}  

\begin{figure*}
\begin{center}
\includegraphics[scale=0.7, angle=0]{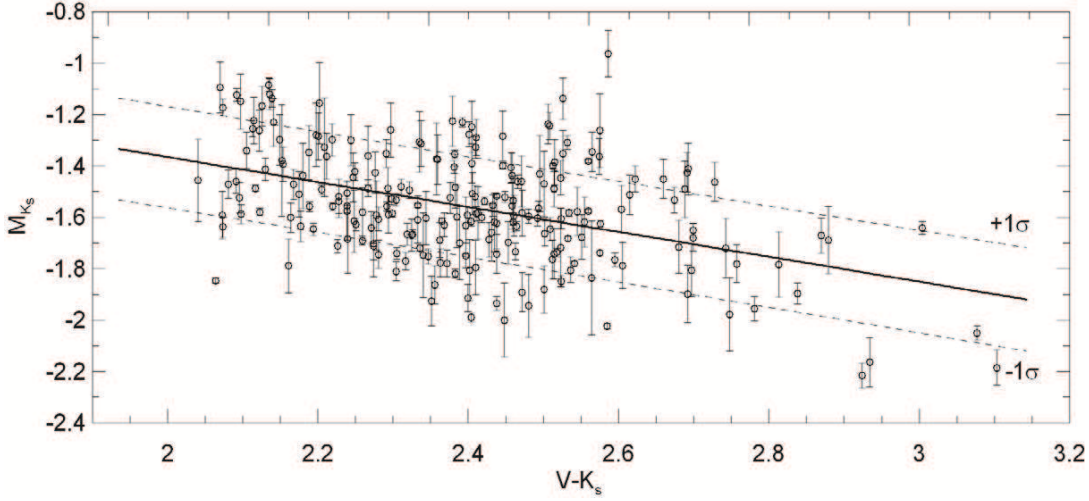}
\caption{Calibration of $M_{K_s}$ absolute magnitude in terms of $V-K_s$.}
\label{histogram}
\end{center}
\end{figure*}  

\begin{figure}
\begin{center}
\includegraphics[scale=0.4, angle=0]{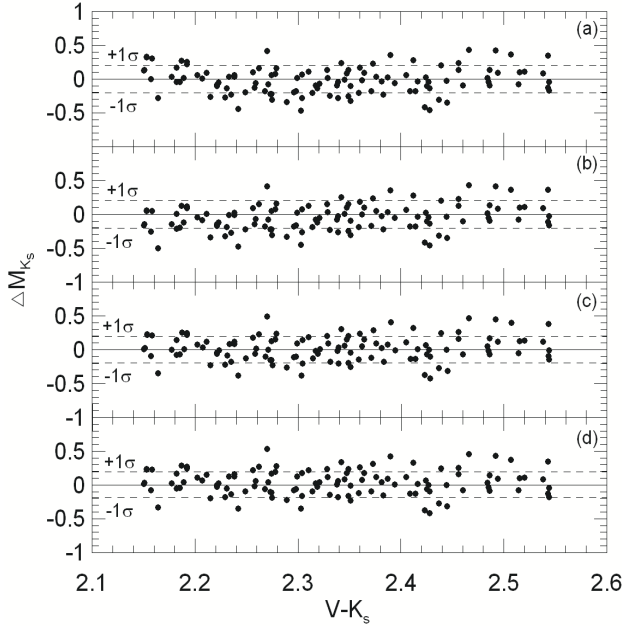}
\caption{Residuals for the data in \citet{Laney12}, panels (a) and (b); and in \citet{Alves00}, panels (c) and (d). Residuals in (a) and (c) are evaluated by the linear equation given in this study, while those in (b) and (d) correspond to the constant absolute magnitude, $M_{K_s}=-1.613$ mag.}
\end{center}
\end{figure}  

\begin{figure}
\begin{center}
\includegraphics[scale=0.4, angle=0]{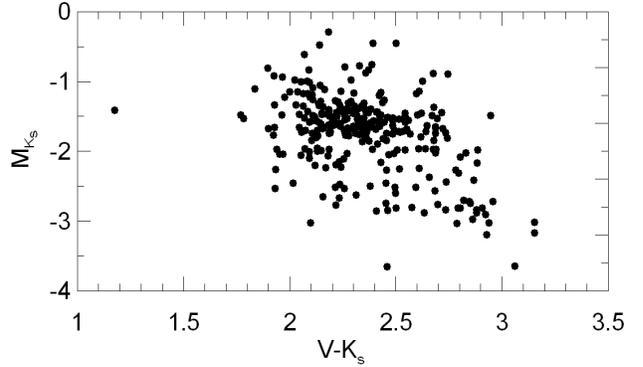}
\caption{$M_{K_s}\times(V-K_s)$ colour absolute magnitude diagram for 282 RC stars in \cite{Alves00}. The absolute magnitudes are estimated by using the corrected parallaxes.}
\end{center}
\end{figure}  

\pagebreak[4]

\begin{table*}
\setlength{\tabcolsep}{5pt}
\center
{\scriptsize
\caption{Data obtained by combination of  the ones in \cite{Laney12} and the $V$ apparent magnitudes in {\em Hipparcos} catalogue. The columns are explained in the text.}
\begin{tabular}{cccccccc|cccccccc}
\hline
 (1) & (2) & (3) &  (4) & (5) & (6) & (7) & (8) & (1) & (2) & (3) & (4) & (5) & (6) & (7) & (8) \\
 ID & Hip & $\pi_0$ &  $[M/H]$ & $V$ & $K_s$ & $V-K_s$ & $M_{K_s}$ & ID & Hip & $\pi_0$ & $[M/H]$ & $V$ & $K_s$ & $V-K_s$ & $M_{K_s}$ \\
    &     &(mas) &   (dex) & (mag) & (mag) & (mag) & (mag)     &    & &(mas) &   (dex) & (mag) & (mag) & (mag) & (mag) \\

\hline
1 &        671 &      10.16 &      -0.07 &       5.99 &      3.653 &      2.337 &     -1.312 &         61 &      23148 &       8.84 &            &       5.84 &      3.744 &      2.096 &     -1.524 \\
         2 &        765 &      22.62 &            &       3.88 &      1.561 &      2.319 &     -1.667 &         62 &      23430 &      13.99 &       0.14 &       5.01 &      2.642 &      2.368 &     -1.629 \\
         3 &        814 &      12.81 &            &       5.29 &      2.968 &      2.322 &     -1.494 &         63 &      23595 &      17.99 &            &       4.55 &      1.769 &      2.781 &     -1.956 \\
         4 &        966 &       8.41 &      -0.12 &       6.53 &      4.151 &      2.379 &     -1.225 &         64 &      25247 &      20.73 &      -0.21 &       4.13 &      1.925 &      2.205 &     -1.492 \\
         5 &       3137 &      10.62 &       0.03 &       6.00 &      3.485 &      2.515 &     -1.385 &         65 &      25532 &       8.34 &            &       6.08 &      3.802 &      2.278 &     -1.592 \\
         6 &       3455 &      13.96 &      -0.06 &       4.77 &      2.466 &      2.304 &     -1.810 &         66 &      26001 &      14.08 &            &       5.34 &      2.930 &      2.410 &     -1.327 \\
         7 &       3456 &       9.63 &      -0.01 &       5.90 &      3.295 &      2.605 &     -1.787 &         67 &      26019 &      12.67 &            &       5.75 &      3.242 &      2.508 &     -1.244 \\
         8 &       3781 &      15.40 &            &       5.09 &      2.013 &      3.077 &     -2.050 &         68 &      26661 &       6.85 &            &       6.77 &      4.673 &      2.097 &     -1.149 \\
         9 &       4257 &       7.61 &      -0.46 &       6.15 &      3.910 &      2.240 &     -1.683 &         69 &      27280 &      10.95 &            &       5.78 &      3.710 &      2.070 &     -1.093 \\
        10 &       4587 &      10.20 &      -0.23 &       5.62 &      3.381 &      2.239 &     -1.576 &         70 &      27369 &       7.21 &            &       6.54 &      4.127 &      2.413 &     -1.583 \\
        11 &       5170 &       8.52 &       0.02 &       6.12 &      3.689 &      2.431 &     -1.659 &         71 &      27530 &      18.45 &            &       4.50 &      2.036 &      2.464 &     -1.634 \\
        12 &       5364 &      26.32 &       0.09 &       3.46 &      0.875 &      2.585 &     -2.023 &         72 &      27549 &       9.10 &            &       5.79 &      3.749 &      2.041 &     -1.456 \\
        13 &       5485 &       8.43 &       0.08 &       6.40 &      4.065 &      2.335 &     -1.306 &         73 &      27621 &      12.72 &            &       5.16 &      2.921 &      2.239 &     -1.556 \\
        14 &       6537 &      28.66 &      -0.10 &       3.60 &      1.178 &      2.422 &     -1.536 &         74 &      27628 &      37.41 &            &       3.12 &      0.560 &      2.560 &     -1.575 \\
        15 &       6592 &      13.84 &      -0.18 &       5.42 &      3.019 &      2.401 &     -1.276 &         75 &      27654 &      28.68 &      -0.63 &       3.76 &      1.314 &      2.446 &     -1.398 \\
        16 &       6732 &      11.69 &       0.00 &       5.50 &      2.998 &      2.502 &     -1.663 &         76 &      27766 &      10.13 &            &       5.62 &      3.223 &      2.397 &     -1.749 \\
        17 &       6868 &       8.73 &            &       6.22 &      3.760 &      2.460 &     -1.535 &         77 &      28011 &      10.31 &            &       5.87 &      3.315 &      2.555 &     -1.619 \\
        18 &       7083 &      22.95 &      -0.28 &       3.93 &      1.638 &      2.292 &     -1.558 &         78 &      28085 &      11.28 &            &       5.95 &      3.376 &      2.574 &     -1.363 \\
        19 &       7271 &       8.02 &      -0.34 &       6.11 &      3.839 &      2.271 &     -1.640 &         79 &      28139 &      10.57 &            &       5.89 &      3.419 &      2.471 &     -1.460 \\
        20 &       7643 &       9.82 &      -0.10 &       5.94 &      3.557 &      2.383 &     -1.482 &         80 &      28988 &       8.87 &            &       6.48 &      3.915 &      2.565 &     -1.345 \\
        21 &       7879 &       8.28 &            &       6.33 &      4.132 &      2.198 &     -1.278 &         81 &      29233 &       6.85 &            &       6.27 &      3.822 &      2.448 &     -2.000 \\
        22 &       7955 &      15.18 &      -0.20 &       5.25 &      2.845 &      2.405 &     -1.248 &         82 &      29294 &       9.89 &            &       5.72 &      3.282 &      2.438 &     -1.742 \\
        23 &       8404 &      10.16 &            &       5.91 &      3.666 &      2.244 &     -1.300 &         83 &      29575 &      10.61 &            &       5.83 &      3.704 &      2.126 &     -1.167 \\
        24 &       8833 &      18.21 &       0.01 &       4.61 &      2.469 &      2.141 &     -1.230 &         84 &      29807 &      17.87 &            &       4.37 &      2.181 &      2.189 &     -1.558 \\
        25 &       8928 &      14.91 &            &       4.68 &      2.486 &      2.194 &     -1.647 &         85 &      29842 &      10.92 &            &       5.54 &      3.004 &      2.536 &     -1.805 \\
        26 &       9440 &      11.08 &      -0.47 &       5.34 &      3.176 &      2.164 &     -1.601 &         86 &      30565 &      12.08 &            &       5.37 &      3.103 &      2.267 &     -1.486 \\
        27 &       9572 &       9.10 &            &       5.87 &      3.695 &      2.175 &     -1.510 &         87 &      30728 &      10.59 &            &       5.55 &      3.188 &      2.362 &     -1.688 \\
        28 &      10234 &       8.40 &            &       5.94 &      3.666 &      2.274 &     -1.712 &         88 &      31061 &       7.44 &            &       6.68 &      4.383 &      2.297 &     -1.259 \\
        29 &      11095 &      10.28 &            &       5.99 &      3.533 &      2.457 &     -1.407 &         89 &      31977 &       7.55 &            &       6.50 &      4.028 &      2.472 &     -1.583 \\
        30 &      11381 &      11.89 &            &       5.89 &      3.199 &      2.691 &     -1.425 &         90 &      32222 &       7.31 &            &       6.36 &      4.101 &      2.259 &     -1.579 \\
        31 &      11524 &       9.09 &       0.04 &       6.11 &      3.652 &      2.458 &     -1.555 &         91 &      34142 &       9.44 &            &       6.09 &      3.410 &      2.680 &     -1.715 \\
        32 &      11757 &      11.49 &            &       5.27 &      2.996 &      2.274 &     -1.702 &         92 &      34270 &       6.18 &            &       6.47 &      4.119 &      2.351 &     -1.926 \\
        33 &      11791 &      12.28 &      -0.05 &       5.36 &      3.067 &      2.293 &     -1.487 &         93 &      34440 &      10.11 &            &       5.47 &      3.114 &      2.356 &     -1.862 \\
        34 &      12148 &      11.02 &            &       5.81 &      3.352 &      2.458 &     -1.437 &         94 &      35044 &      12.00 &            &       5.58 &      2.823 &      2.757 &     -1.781 \\
        35 &      12486 &      21.65 &      -0.33 &       4.11 &      1.706 &      2.404 &     -1.617 &         95 &      36444 &       9.06 &            &       5.87 &      3.545 &      2.325 &     -1.669 \\
        36 &      12608 &       7.91 &       0.01 &       5.99 &      3.917 &      2.073 &     -1.592 &         96 &      36732 &      11.96 &            &       5.64 &      3.126 &      2.514 &     -1.486 \\
        37 &      13147 &      18.89 &      -0.34 &       4.45 &      2.139 &      2.311 &     -1.480 &         97 &      37202 &       8.03 &            &       6.20 &      3.867 &      2.333 &     -1.609 \\
        38 &      13288 &      17.45 &      -0.04 &       4.76 &      2.668 &      2.092 &     -1.123 &         98 &      37447 &      22.07 &       0.01 &       3.94 &      1.614 &      2.326 &     -1.667 \\
        39 &      13701 &      23.89 &      -0.11 &       3.89 &      1.372 &      2.518 &     -1.737 &         99 &      37504 &      23.13 &            &       3.93 &      1.565 &      2.365 &     -1.614 \\
        40 &      14060 &      10.07 &            &       5.75 &      3.353 &      2.397 &     -1.632 &        100 &      37590 &      10.33 &            &       5.64 &      3.412 &      2.228 &     -1.518 \\
        41 &      14168 &      12.32 &            &       5.32 &      3.168 &      2.152 &     -1.379 &        101 &      37664 &      15.03 &            &       5.12 &      2.654 &      2.466 &     -1.461 \\
        42 &      14838 &      19.22 &       0.09 &       4.35 &      2.052 &      2.298 &     -1.529 &        102 &      37901 &      12.18 &            &       5.49 &      2.387 &      3.103 &     -2.185 \\
        43 &      16290 &      10.17 &            &       5.68 &      3.550 &      2.130 &     -1.413 &        103 &      38211 &      12.81 &            &       5.17 &      2.246 &      2.924 &     -2.216 \\
        44 &      17086 &       7.49 &            &       6.22 &      3.880 &      2.340 &     -1.748 &        104 &      38375 &      10.67 &            &       5.62 &      3.466 &      2.154 &     -1.393 \\
        45 &      17351 &      17.70 &            &       4.59 &      1.995 &      2.595 &     -1.765 &        105 &      40084 &      18.46 &       0.09 &       4.72 &      2.585 &      2.135 &     -1.084 \\
        46 &      17595 &       9.73 &            &       5.91 &      3.634 &      2.276 &     -1.425 &        106 &      40107 &      11.22 &            &       5.36 &      3.279 &      2.081 &     -1.471 \\
        47 &      17738 &      12.13 &      -0.31 &       5.52 &      3.229 &      2.291 &     -1.351 &        107 &      40888 &      21.00 &            &       4.34 &      1.764 &      2.576 &     -1.625 \\
        48 &      18199 &       9.29 &       0.13 &       5.93 &      3.182 &      2.748 &     -1.978 &        108 &      40990 &      10.50 &            &       5.71 &      3.275 &      2.435 &     -1.619 \\
        49 &      18401 &       9.66 &      -0.10 &       6.14 &      3.645 &      2.495 &     -1.430 &        109 &      41191 &      13.18 &            &       5.67 &      3.164 &      2.506 &     -1.236 \\
        50 &      18554 &       5.62 &            &       6.94 &      4.551 &      2.389 &     -1.700 &        110 &      41312 &      30.33 &            &       3.77 &      1.210 &      2.560 &     -1.381 \\
        51 &      18635 &       6.24 &            &       6.83 &      4.307 &      2.523 &     -1.717 &        111 &      41321 &       8.90 &            &       5.95 &      3.782 &      2.168 &     -1.471 \\
        52 &      19511 &      11.40 &            &       5.70 &      3.341 &      2.359 &     -1.375 &        112 &      41395 &      11.86 &            &       5.52 &      2.823 &      2.697 &     -1.806 \\
        53 &      19747 &      28.36 &      -0.02 &       3.85 &      1.337 &      2.513 &     -1.399 &        113 &      41907 &       8.25 &            &       6.11 &      3.882 &      2.228 &     -1.536 \\
        54 &      19805 &      12.41 &            &       5.45 &      2.926 &      2.524 &     -1.605 &        114 &      41939 &       7.87 &            &       6.36 &      4.172 &      2.188 &     -1.348 \\
        55 &      20161 &      12.94 &            &       5.33 &      2.838 &      2.492 &     -1.603 &        115 &      42134 &      14.69 &            &       4.84 &      2.559 &      2.281 &     -1.606 \\
        56 &      20825 &       9.11 &            &       5.74 &      3.459 &      2.281 &     -1.744 &        116 &      42662 &      15.40 &            &       4.87 &      2.471 &      2.399 &     -1.591 \\
        57 &      20877 &      17.47 &      -0.05 &       4.96 &      2.338 &      2.622 &     -1.450 &        117 &      42717 &      10.03 &            &       6.26 &      3.532 &      2.728 &     -1.462 \\
        58 &      21594 &      29.69 &       0.01 &       3.86 &      1.328 &      2.532 &     -1.309 &        118 &      42911 &      24.98 &      -0.01 &       3.94 &      1.491 &      2.449 &     -1.521 \\
        59 &      22081 &       6.72 &            &       6.46 &      4.058 &      2.402 &     -1.805 &        119 &      42915 &       8.40 &            &       6.66 &      3.967 &      2.693 &     -1.411 \\
        60 &      22479 &      13.83 &       0.05 &       5.03 &      2.791 &      2.239 &     -1.505 &        120 &      43026 &      10.80 &            &       5.70 &      3.239 &      2.461 &     -1.594 \\
\hline
\end{tabular} 
}
\end{table*}

\begin{table*}
\contcaption{Continued}
\setlength{\tabcolsep}{5pt}
{\scriptsize
\begin{tabular}{cccccccc|cccccccc}
\hline
 (1) & (2) & (3) &  (4) & (5) & (6) & (7) & (8) & (1) & (2) & (3) & (4) & (5) & (6) & (7) & (8) \\
 ID & Hip & $\pi$ &  $[M/H]$ & $V$ & $K_s$ & $V-K_s$ & $M_{K_s}$ & ID & Hip & $\pi$ & $[M/H]$ & $V$ & $K_s$ & $V-K_s$ & $M_{K_s}$ \\
    &     &(mas) &   (dex) & (mag) & (mag) & (mag) & (mag)     &    & &(mas) &   (dex) & (mag) & (mag) & (mag) & (mag) \\
\hline
       121 &      43580 &       9.62 &            &       6.12 &      3.516 &      2.604 &     -1.568 &        173 &      70306 &      20.72 &            &       4.78 &      1.776 &      3.004 &     -1.642 \\
       122 &      45166 &       8.39 &            &       6.13 &      3.618 &      2.512 &     -1.763 &        174 &      72471 &      12.10 &            &       6.21 &      3.624 &      2.586 &     -0.962 \\
       123 &      45439 &      14.52 &            &       4.92 &      2.457 &      2.463 &     -1.733 &        175 &      73620 &      16.69 &      -0.10 &       4.39 &      1.952 &      2.438 &     -1.935 \\
       124 &      45542 &       6.13 &            &       7.11 &      4.908 &      2.202 &     -1.155 &        176 &      74395 &      27.80 &            &       3.41 &      1.293 &      2.117 &     -1.487 \\
       125 &      45796 &       7.34 &            &       6.70 &      3.887 &      2.813 &     -1.784 &        177 &      75119 &      13.63 &      -0.02 &       5.35 &      2.607 &      2.743 &     -1.720 \\
       126 &      45811 &      14.66 &       0.05 &       4.80 &      2.709 &      2.091 &     -1.460 &        178 &      75127 &      10.79 &       0.06 &       5.54 &      3.246 &      2.294 &     -1.589 \\
       127 &      45856 &      13.96 &            &       4.79 &      2.564 &      2.226 &     -1.712 &        179 &      76333 &      19.99 &      -0.30 &       3.91 &      1.506 &      2.404 &     -1.990 \\
       128 &      46026 &      16.98 &      -0.06 &       4.71 &      2.597 &      2.113 &     -1.254 &        180 &      76532 &      11.87 &       0.24 &       5.79 &      3.344 &      2.446 &     -1.284 \\
       129 &      46371 &      20.83 &       0.11 &       4.72 &      2.145 &      2.575 &     -1.262 &        181 &      76664 &      10.09 &      -0.04 &       6.19 &      3.530 &      2.660 &     -1.450 \\
       130 &      46736 &      11.69 &            &       5.86 &      2.990 &      2.870 &     -1.671 &        182 &      77070 &      44.10 &       0.16 &       2.63 &      0.097 &      2.533 &     -1.681 \\
       131 &      46771 &      15.13 &      -0.05 &       4.99 &      2.581 &      2.409 &     -1.520 &        183 &      77578 &      11.93 &      -0.18 &       5.21 &      2.838 &      2.372 &     -1.779 \\
       132 &      46869 &       7.99 &            &       6.12 &      3.872 &      2.248 &     -1.615 &        184 &      77853 &      19.36 &      -0.21 &       4.13 &      1.747 &      2.383 &     -1.819 \\
       133 &      47172 &       8.78 &            &       6.18 &      3.774 &      2.406 &     -1.508 &        185 &      78639 &      14.86 &      -0.05 &       4.65 &      2.552 &      2.098 &     -1.588 \\
       134 &      47205 &      13.28 &            &       5.00 &      2.590 &      2.410 &     -1.794 &        186 &      78650 &      15.71 &      -0.01 &       4.96 &      2.122 &      2.838 &     -1.897 \\
       135 &      48119 &       9.84 &            &       6.05 &      3.914 &      2.136 &     -1.121 &        187 &      78685 &       9.04 &       0.01 &       6.07 &      3.921 &      2.149 &     -1.298 \\
       136 &      48806 &       6.86 &            &       6.59 &      4.076 &      2.514 &     -1.742 &        188 &      79666 &       9.38 &       0.13 &       5.72 &      3.248 &      2.472 &     -1.891 \\
       137 &      49418 &       7.13 &            &       6.27 &      3.790 &      2.480 &     -1.944 &        189 &      79882 &      30.64 &      -0.10 &       3.23 &      1.010 &      2.220 &     -1.558 \\
       138 &      49477 &       7.61 &            &       6.50 &      4.233 &      2.267 &     -1.360 &        190 &      80000 &      25.33 &       0.23 &       4.01 &      1.628 &      2.382 &     -1.354 \\
       139 &      49841 &      28.98 &       0.17 &       3.61 &      1.391 &      2.219 &     -1.298 &        191 &      80343 &      16.35 &      -0.13 &       4.48 &      2.163 &      2.317 &     -1.770 \\
       140 &      50234 &       9.31 &            &       6.17 &      3.765 &      2.405 &     -1.391 &        192 &      81852 &      20.78 &      -0.05 &       4.23 &      1.812 &      2.418 &     -1.600 \\
       141 &      50799 &      16.00 &            &       4.82 &      2.360 &      2.460 &     -1.619 &        193 &      82396 &      51.19 &      -0.05 &       2.29 &     -0.285 &      2.575 &     -1.739 \\
       142 &      51077 &      10.20 &            &       6.13 &      3.604 &      2.526 &     -1.353 &        194 &      83000 &      35.66 &       0.09 &       3.19 &      0.656 &      2.534 &     -1.583 \\
       143 &      52085 &      15.49 &      -0.10 &       4.91 &      2.788 &      2.122 &     -1.262 &        195 &      86170 &      19.67 &      -0.25 &       4.26 &      1.696 &      2.564 &     -1.835 \\
       144 &      52660 &      10.04 &            &       6.38 &      3.854 &      2.526 &     -1.137 &        196 &      86391 &       8.85 &       0.14 &       6.25 &      3.892 &      2.358 &     -1.373 \\
       145 &      52689 &      11.35 &            &       5.49 &      3.385 &      2.105 &     -1.340 &        197 &      86742 &      39.85 &       0.14 &       2.76 &      0.219 &      2.541 &     -1.779 \\
       146 &      52948 &       9.57 &            &       5.85 &      3.397 &      2.453 &     -1.698 &        198 &      88635 &      33.67 &      -0.24 &       2.98 &      0.644 &      2.336 &     -1.720 \\
       147 &      53273 &      10.43 &            &       5.45 &      3.273 &      2.177 &     -1.635 &        199 &      89153 &      12.74 &      -0.06 &       4.96 &      2.560 &      2.400 &     -1.914 \\
       148 &      53394 &      10.70 &            &       5.93 &      3.407 &      2.523 &     -1.446 &        200 &      89587 &       9.67 &      -0.60 &       5.99 &      3.789 &      2.201 &     -1.284 \\
       149 &      53502 &      17.16 &            &       4.60 &      2.296 &      2.304 &     -1.532 &        201 &      90496 &      41.72 &      -0.07 &       2.82 &      0.382 &      2.438 &     -1.516 \\
       150 &      54264 &       8.03 &            &       6.28 &      4.033 &      2.247 &     -1.444 &        202 &      90568 &      25.84 &      -0.20 &       4.10 &      1.708 &      2.392 &     -1.230 \\
       151 &      54291 &       8.54 &            &       6.31 &      3.765 &      2.545 &     -1.577 &        203 &      93498 &      12.66 &       0.36 &       5.63 &      2.956 &      2.674 &     -1.532 \\
       152 &      55249 &      11.08 &            &       5.90 &      3.489 &      2.411 &     -1.288 &        204 &      93683 &      22.96 &      -0.01 &       3.76 &      1.455 &      2.305 &     -1.740 \\
       153 &      56287 &      10.64 &            &       5.89 &      3.376 &      2.514 &     -1.490 &        205 &      94005 &      18.27 &       0.01 &       4.57 &      2.137 &      2.433 &     -1.554 \\
       154 &      56343 &      25.16 &       0.08 &       3.54 &      1.417 &      2.123 &     -1.579 &        206 &      98575 &       9.29 &       0.04 &       6.01 &      3.798 &      2.212 &     -1.362 \\
       155 &      56656 &      13.95 &            &       5.14 &      2.631 &      2.509 &     -1.646 &        207 &      98624 &      13.80 &       0.11 &       5.32 &      2.621 &      2.699 &     -1.679 \\
       156 &      56996 &       9.05 &            &       6.32 &      3.705 &      2.615 &     -1.512 &        208 &      99570 &       9.99 &       0.22 &       6.20 &      3.512 &      2.688 &     -1.490 \\
       157 &      57791 &      10.94 &            &       5.62 &      3.182 &      2.438 &     -1.623 &        209 &     101772 &      33.17 &      -0.13 &       3.11 &      0.811 &      2.299 &     -1.585 \\
       158 &      58697 &       8.79 &            &       6.05 &      3.116 &      2.934 &     -2.164 &        210 &     103738 &      14.24 &      -0.11 &       4.67 &      2.596 &      2.074 &     -1.637 \\
       159 &      58706 &       7.53 &            &       6.41 &      3.718 &      2.692 &     -1.898 &        211 &     105425 &       9.08 &            &       6.40 &      3.521 &      2.879 &     -1.688 \\
       160 &      58948 &      19.98 &      -0.39 &       4.12 &      1.869 &      2.251 &     -1.628 &        212 &     106039 &      19.06 &      -0.09 &       4.50 &      2.426 &      2.074 &     -1.173 \\
       161 &      59785 &       8.12 &            &       6.24 &      3.855 &      2.385 &     -1.597 &        213 &     111600 &       9.12 &      -0.03 &       5.82 &      3.319 &      2.501 &     -1.881 \\
       162 &      61181 &       9.38 &            &       5.88 &      3.452 &      2.428 &     -1.687 &        214 &     112127 &       9.21 &            &       6.06 &      3.851 &      2.209 &     -1.327 \\
       163 &      62012 &      17.11 &            &       4.66 &      2.249 &      2.411 &     -1.585 &        215 &     112203 &      14.16 &      -0.20 &       4.84 &      2.493 &      2.347 &     -1.752 \\
       164 &      63608 &      29.76 &       0.27 &       2.85 &      0.786 &      2.064 &     -1.846 &        216 &     113246 &      21.16 &      -0.20 &       4.20 &      1.951 &      2.249 &     -1.421 \\
       165 &      65468 &      14.75 &            &       5.04 &      2.560 &      2.480 &     -1.596 &        217 &     114119 &      15.08 &      -0.02 &       4.48 &      2.319 &      2.161 &     -1.789 \\
       166 &      66936 &      13.68 &       0.11 &       5.35 &      2.849 &      2.501 &     -1.470 &        218 &     114855 &      21.77 &      -0.01 &       4.24 &      1.746 &      2.494 &     -1.565 \\
       167 &      67494 &      13.35 &       0.09 &       4.96 &      2.597 &      2.363 &     -1.776 &        219 &     114971 &      23.64 &      -0.52 &       3.70 &      1.441 &      2.259 &     -1.691 \\
       168 &      68079 &      10.25 &            &       5.82 &      3.269 &      2.551 &     -1.677 &        220 &     115102 &      17.90 &      -0.08 &       4.41 &      1.886 &      2.524 &     -1.849 \\
       169 &      68933 &      55.45 &            &       2.06 &     -0.273 &      2.333 &     -1.554 &        221 &     115620 &      11.23 &       0.08 &       5.60 &      3.224 &      2.376 &     -1.524 \\
       170 &      69191 &      17.88 &            &       4.74 &      2.601 &      2.139 &     -1.137 &        222 &     115830 &      21.96 &       0.03 &       4.27 &      1.889 &      2.381 &     -1.403 \\
       171 &      69612 &      12.31 &      -0.14 &       5.29 &      2.946 &      2.344 &     -1.603 &        223 &     115919 &      18.65 &       0.07 &       4.54 &      2.425 &      2.115 &     -1.222 \\
       172 &      70027 &      17.44 &       0.12 &       4.84 &      2.141 &      2.699 &     -1.651 &        224 &     116853 &       9.34 &       0.10 &       5.89 &      3.710 &      2.180 &     -1.438 \\
\hline
\end{tabular} 
}
\end{table*}

\begin{table}
\setlength{\tabcolsep}{2pt}
{\small
\center
\caption{Mean residuals and the corresponding standard deviations for the data in \citet{Laney12} and \citet{Alves00}. The statistics in the second and third columns are evaluated by the linear equation in Eq. (4), while those in the fourth and fifth columns correspond to the ones where the $M_{K_s}$ absolute magnitude is adopted as a constant value, $M_{K_s}=-1.613$ mag.}
\begin{tabular}{lcccc}
\hline
& \multicolumn{2}{c}{Linear Equation} & \multicolumn{2}{c}{$M_{K_s}=-1.613$}\\ 
\hline
Study            & $<\Delta M_{K_s}>$ & $\sigma$ & $<\Delta M_{K_s}>$ & $\sigma$\\ 
\hline
\citet{Laney12} &--0.001             &   0.195  &--0.053 & 0.218 \\
\citet{Alves00} & +0.209             &   0.524  & +0.133 & 0.571 \\
\hline
\end{tabular}
}  
\end{table}

\begin{table*}
\setlength{\tabcolsep}{3pt}
\center
{\tiny
\caption{Data taken from \cite{Alves00}. The columns are explained in the text.}
\begin{tabular}{ccccccccc|ccccccccc}
\hline
(1) & (2) &  (3) & (4) & (5) & (6) & (7) & (8) & (9) & (1) & (2) &  (3) & (4) & (5) & (6) & (7) & (8) & (9)\\
ID & Hip &    $\pi_0$ & $[M/H]$ & $V$ & $K_s$ & $V-K_s$ &$M_{K_s}$ & Quality & ID & Hip & $\pi_0$ & $[M/H]$ & $V$ & $K_s$ &$V-K_s$ & $M_{K_s}$ &    Quality \\
& &    (mas)  & (dex)  &(mag)& (mag) & (mag)   & (mag)    &         &    &     & (mas)   & (dex)   & (mag) & (mag) &(mag)   & (mag)     & \\
\hline
         1 &        443 &      25.28 &      -0.31 &      4.613 &      1.989 &      2.624 &     -0.997 &        DCD &         76 &      30277 &      13.86 &      -0.31 &      3.852 &      1.836 &      2.016 &     -2.455 &        DDD \\
         2 &        476 &       8.71 &      -0.22 &      5.550 &      3.767 &      1.783 &     -1.533 &        DDD &         77 &      32249 &      11.27 &      -0.26 &      4.493 &      1.861 &      2.632 &     -2.879 &        DCD \\
         3 &        729 &      11.40 &      -0.01 &      5.570 &      3.321 &      2.249 &     -1.394 &        DDD &         78 &      33449 &      18.28 &       0.05 &      4.350 &            &            &            &        DCX \\
         4 &        873 &      13.29 &      -0.02 &      5.841 &      3.772 &      2.069 &     -0.610 &        DDD &         79 &      36046 &      27.09 &      -0.17 &      3.777 &      1.562 &      2.215 &     -1.274 &        DDC \\
         5 &       1708 &       9.80 &       0.00 &      5.176 &      3.045 &      2.131 &     -1.999 &        DDD &         80 &      37447 &      22.06 &      -0.11 &      3.942 &      1.643 &      2.299 &     -1.639 &        DDD \\
         6 &       2568 &      12.47 &      -0.08 &      5.377 &      2.950 &      2.427 &     -1.571 &        DCD &         81 &      37740 &      23.06 &      -0.16 &      3.573 &      1.527 &      2.046 &     -1.659 &        DCD \\
         7 &       3031 &      19.90 &      -0.64 &      4.342 &      2.074 &      2.268 &     -1.432 &        DCD &         82 &      37826 &      96.54 &      -0.07 &      1.158 &     -0.936 &      2.094 &     -1.012 &        DCC \\
         8 &       3092 &      30.91 &       0.04 &      3.269 &      0.467 &      2.802 &     -2.083 &        DCC &         83 &      39079 &      11.85 &      -0.24 &      4.930 &      2.192 &      2.738 &     -2.439 &        DDD \\
         9 &       3231 &       9.15 &      -0.26 &      5.302 &      3.231 &      2.071 &     -1.962 &        DCD &         84 &      39424 &      12.47 &       0.03 &      4.942 &      2.476 &      2.466 &     -2.045 &        DCD \\
        10 &       3419 &      33.86 &      -0.09 &      2.040 &     -0.274 &      2.314 &     -2.626 &        CDD &         85 &      40084 &      18.45 &      -0.03 &      4.723 &      2.670 &      2.053 &     -1.000 &        DDD \\
        11 &       3455 &      13.94 &      -0.16 &      4.767 &      2.575 &      2.192 &     -1.704 &        DCC &         86 &      41325 &       8.21 &      -0.34 &      5.130 &      2.915 &      2.215 &     -2.513 &        DCD \\
        12 &       4422 &      16.31 &      -0.51 &      4.619 &      2.468 &      2.151 &     -1.470 &        DCD &         87 &      42483 &      13.33 &      -0.49 &      4.864 &      2.900 &      1.964 &     -1.476 &        DDD \\
        13 &       4463 &      14.18 &      -0.54 &      4.402 &      2.254 &      2.148 &     -1.988 &        DDD &         88 &      42527 &      12.72 &      -0.26 &      4.586 &      1.734 &      2.852 &     -2.744 &        DCD \\
        14 &       4587 &      10.09 &      -0.37 &      5.615 &      3.413 &      2.202 &     -1.568 &        DCC &         89 &      42662 &      15.39 &      -0.01 &      4.868 &      2.637 &      2.231 &     -1.427 &        DDD \\
        15 &       4906 &      17.93 &      -0.39 &      4.270 &      2.121 &      2.149 &     -1.611 &        DDD &         90 &      42911 &      24.97 &      -0.13 &      3.938 &      1.567 &      2.371 &     -1.446 &        DCC \\
        16 &       5364 &      26.32 &      -0.03 &      3.465 &      0.922 &      2.543 &     -1.977 &        DDD &         91 &      43409 &      15.72 &      -0.11 &      4.020 &      1.140 &      2.880 &     -2.878 &        DCD \\
        17 &       5586 &      19.31 &      -0.04 &      4.507 &      2.148 &      2.359 &     -1.423 &        DCD &         92 &      43813 &      19.50 &      -0.21 &      3.106 &      0.697 &      2.409 &     -2.853 &        DCC \\
        18 &       6411 &      15.19 &       0.03 &      4.871 &      2.554 &      2.317 &     -1.538 &        DDD &         93 &      45751 &       9.55 &      -0.20 &      4.773 &      2.840 &      1.933 &     -2.260 &        DDD \\
        19 &       6537 &      28.65 &      -0.22 &      3.603 &      1.289 &      2.314 &     -1.425 &        DDD &         94 &      45811 &      14.63 &      -0.07 &      3.936 &      2.761 &      1.175 &     -1.413 &        DCD \\
        20 &       6692 &      16.71 &      -0.13 &      4.718 &      2.324 &      2.394 &     -1.561 &        DCD &         95 &      46026 &      16.97 &      -0.18 &      4.707 &      2.707 &      2.000 &     -1.145 &        DDD \\
        21 &       6732 &      11.67 &      -0.12 &      5.503 &      3.087 &      2.416 &     -1.578 &        DDD &         96 &      46146 &      16.19 &       0.01 &      4.471 &      1.688 &      2.783 &     -2.266 &        DCC \\
        22 &       6999 &      11.06 &      -0.16 &      5.268 &      3.115 &      2.153 &     -1.666 &        DCD &         97 &      46371 &      20.46 &      -0.01 &      4.719 &      2.291 &      2.428 &     -1.154 &        DDD \\
        23 &       7294 &      15.63 &      -0.36 &      4.675 &      2.311 &      2.364 &     -1.719 &        CCD &         98 &      46771 &      15.10 &      -0.17 &      4.989 &      2.716 &      2.273 &     -1.389 &        DDD \\
        24 &       7607 &      18.40 &       0.00 &      3.586 &      0.649 &      2.937 &     -3.027 &        DBC &         99 &      46952 &      17.62 &      -0.15 &      4.538 &            &            &            &        DDX \\
        25 &       7719 &      12.94 &      -0.21 &      5.015 &      2.969 &      2.046 &     -1.471 &        DCD &        100 &      47029 &      15.13 &      -0.18 &      4.806 &      2.629 &      2.177 &     -1.472 &        DCD \\
        26 &       7955 &      15.05 &      -0.29 &      5.253 &      2.950 &      2.303 &     -1.162 &        DCD &        101 &      48455 &      26.28 &       0.17 &      3.878 &      1.364 &      2.514 &     -1.538 &        DCD \\
        27 &       8198 &      11.51 &      -0.11 &      4.257 &      2.025 &      2.232 &     -2.670 &        DDD &        102 &      48559 &       9.49 &      -0.03 &      4.866 &      2.078 &      2.788 &     -3.036 &        DDD \\
        28 &       8833 &      18.08 &      -0.11 &      4.606 &      2.541 &      2.065 &     -1.173 &        DDD &        103 &      49841 &      28.88 &       0.05 &      3.610 &      1.525 &      2.085 &     -1.172 &        DDD \\
        29 &       9440 &      11.05 &      -0.59 &      5.343 &      3.218 &      2.125 &     -1.565 &        DDD &        104 &      51233 &      21.14 &       0.00 &      4.196 &      2.061 &      2.135 &     -1.313 &        CCD \\
        30 &       9763 &       8.12 &      -0.10 &      5.218 &      2.982 &      2.236 &     -2.470 &        DCD &        105 &      51775 &      12.24 &      -0.25 &      5.069 &      2.977 &      2.092 &     -1.584 &        DDD \\
        31 &       9884 &      49.55 &      -0.25 &      2.012 &     -0.783 &      2.795 &     -2.308 &        DCC &        106 &      52085 &      15.41 &      -0.22 &      4.910 &      2.800 &      2.110 &     -1.261 &        DDD \\
        32 &      10642 &       9.73 &      -0.10 &      5.507 &      3.320 &      2.187 &     -1.739 &        DCD &        107 &      52353 &      14.36 &       0.06 &      5.119 &      2.234 &      2.885 &     -1.980 &        CCD \\
        33 &      11791 &      12.21 &      -0.17 &      5.364 &      3.207 &      2.157 &     -1.359 &        DCD &        108 &      52943 &      22.68 &      -0.30 &      3.109 &      0.248 &      2.861 &     -2.974 &        DDD \\
        34 &      13061 &      19.00 &      -0.02 &      4.524 &      2.099 &      2.425 &     -1.507 &        DCD &        109 &      53229 &      34.37 &      -0.20 &      3.789 &      1.530 &      2.259 &     -0.789 &        FCF \\
        35 &      13147 &      18.88 &      -0.47 &      4.449 &      2.098 &      2.351 &     -1.522 &        DCD &        110 &      53426 &      13.75 &      -0.26 &      5.024 &      2.334 &      2.690 &     -1.974 &        DDD \\
        36 &      13288 &      17.44 &      -0.17 &      4.758 &      2.483 &      2.275 &     -1.309 &        CCD &        111 &      53740 &      20.48 &      -0.22 &      4.080 &      1.733 &      2.347 &     -1.710 &        DDD \\
        37 &      13701 &      23.88 &      -0.23 &      3.895 &      1.471 &      2.424 &     -1.639 &        DDD &        112 &      53807 &       9.03 &      -0.28 &      4.838 &      2.377 &      2.461 &     -2.845 &        DDD \\
        38 &      14382 &      15.43 &      -0.17 &      4.774 &      2.398 &      2.376 &     -1.660 &        DCD &        113 &      54539 &      22.57 &      -0.13 &      3.003 &      0.429 &      2.574 &     -2.803 &        CCC \\
        39 &      14668 &      28.92 &       0.04 &      3.786 &      1.242 &      2.544 &     -1.452 &        DBC &        114 &      56343 &      25.16 &      -0.04 &      3.540 &      1.471 &      2.069 &     -1.525 &        DDD \\
        40 &      14817 &      11.30 &      -0.10 &      4.615 &      2.235 &      2.380 &     -2.500 &        DCD &        115 &      56647 &      17.96 &      -0.34 &      4.304 &      2.184 &      2.120 &     -1.544 &        DDD \\
        41 &      14838 &      19.21 &      -0.03 &      4.354 &      2.169 &      2.185 &     -1.413 &        DCD &        116 &      57283 &       9.09 &      -0.11 &      4.715 &      2.558 &      2.157 &     -2.649 &        DCD \\
        42 &      15382 &      12.33 &      -0.07 &      4.864 &      2.741 &      2.123 &     -1.804 &        DDD &        117 &      57399 &      17.75 &      -0.44 &      3.686 &      0.988 &      2.698 &     -2.766 &        DCC \\
        43 &      15383 &      14.70 &      -0.10 &      5.624 &      2.678 &      2.946 &     -1.485 &        CCD &        118 &      58948 &      19.97 &      -0.51 &      4.123 &      2.014 &      2.109 &     -1.484 &        DDD \\
        44 &      15861 &      15.65 &       0.00 &      5.498 &      2.888 &      2.610 &     -1.139 &        DDD &        119 &      59856 &      10.25 &      -0.29 &      4.987 &      2.106 &      2.881 &     -2.840 &        DCD \\
        45 &      15900 &      10.93 &      -0.15 &      3.610 &      1.152 &      2.458 &     -3.655 &        DCC &        120 &      60172 &      10.57 &      -0.48 &      4.970 &      2.065 &      2.905 &     -2.815 &        DDD \\
        46 &      16780 &       8.55 &      -0.30 &      5.563 &      3.663 &      1.900 &     -1.677 &        DDC &        121 &      60646 &      12.45 &      -0.11 &      5.014 &      2.856 &      2.158 &     -1.668 &        DDD \\
        47 &      17738 &      12.10 &      -0.43 &      5.523 &      3.294 &      2.229 &     -1.292 &        DDD &        122 &      61359 &      22.38 &      -0.11 &      2.651 &      0.719 &      1.932 &     -2.532 &        DDD \\
        48 &      19038 &      17.42 &       0.01 &      4.361 &      2.033 &      2.328 &     -1.762 &        DDD &        123 &      63608 &      29.76 &       0.15 &      2.849 &      0.664 &      2.185 &     -1.968 &        DDD \\
        49 &      19388 &      11.44 &      -0.02 &      5.508 &      3.170 &      2.338 &     -1.538 &        DDD &        124 &      64078 &      10.59 &      -0.04 &      5.149 &      2.720 &      2.429 &     -2.156 &        DDD \\
        50 &      19483 &       9.22 &      -0.10 &      5.445 &      3.516 &      1.929 &     -1.660 &        DCD &        125 &      64166 &      14.02 &      -0.19 &      4.942 &      2.735 &      2.207 &     -1.531 &        DDD \\
        51 &      20205 &      20.17 &      -0.02 &      3.649 &      1.518 &      2.131 &     -1.958 &        DCC &        126 &      64803 &      12.64 &      -0.15 &      5.095 &      2.975 &      2.120 &     -1.516 &        DDD \\
        52 &      20266 &      10.16 &      -0.18 &      5.256 &      3.486 &      1.770 &     -1.480 &        DDD &        127 &      64962 &      24.37 &      -0.12 &      2.993 &      1.024 &      1.969 &     -2.042 &        DDD \\
        53 &      20455 &      20.90 &       0.00 &      3.771 &      1.643 &      2.128 &     -1.756 &        DCD &        128 &      65535 &      15.95 &      -0.25 &      5.114 &      2.513 &      2.601 &     -1.473 &        DCD \\
        54 &      20877 &      17.43 &      -0.17 &      4.964 &      2.344 &      2.620 &     -1.450 &        DDD &        129 &      65639 &      11.62 &       0.02 &      4.757 &      2.524 &      2.233 &     -2.150 &        DDD \\
        55 &      20885 &      21.12 &       0.04 &      3.836 &      1.644 &      2.192 &     -1.733 &        DCD &        130 &      66098 &      13.84 &      -0.40 &      5.210 &      3.105 &      2.105 &     -1.189 &        DDD \\
        56 &      20889 &      22.23 &       0.04 &      3.525 &      1.422 &      2.103 &     -1.843 &        DDD &        131 &      66936 &      13.49 &      -0.01 &      5.352 &      2.793 &      2.559 &     -1.557 &        DCD \\
        57 &      21248 &      25.66 &      -0.34 &      4.493 &      2.122 &      2.371 &     -0.832 &        DCD &        132 &      67494 &      13.32 &      -0.03 &      4.959 &      2.610 &      2.349 &     -1.767 &        DDC \\
        58 &      21393 &      15.24 &      -0.09 &      3.808 &      1.552 &      2.256 &     -2.533 &        DDD &        133 &      68895 &      32.30 &      -0.16 &      3.247 &      0.753 &      2.494 &     -1.701 &        DDD \\
        59 &      21594 &      29.67 &      -0.11 &      3.864 &      1.441 &      2.423 &     -1.197 &        DDD &        134 &      69415 &      14.96 &      -0.14 &      5.075 &      2.388 &      2.687 &     -1.737 &        DDD \\
        60 &      21685 &      16.60 &      -0.24 &      5.457 &      3.127 &      2.330 &     -0.772 &        DDD &        135 &      69612 &      12.20 &      -0.26 &      5.294 &      2.949 &      2.345 &     -1.619 &        DDD \\
        61 &      22479 &      13.80 &      -0.07 &      5.026 &      2.803 &      2.223 &     -1.498 &        DCD &        136 &      69879 &      13.93 &      -0.13 &      4.796 &      2.384 &      2.412 &     -1.896 &        DDD \\
        62 &      22957 &      17.53 &      -0.26 &      4.063 &      1.407 &      2.656 &     -2.374 &        DCC &        137 &      70012 &      12.44 &      -0.22 &      5.138 &      2.834 &      2.304 &     -1.692 &        DDD \\
        63 &      23430 &      13.96 &       0.02 &      5.014 &      2.743 &      2.271 &     -1.533 &        DDD &        138 &      70027 &      17.43 &       0.00 &      4.838 &      2.116 &      2.722 &     -1.678 &        DCD \\
        64 &      24822 &      13.16 &      -0.10 &      4.957 &      2.816 &      2.141 &     -1.588 &        DDD &        139 &      71053 &      20.36 &      -0.17 &      3.574 &      0.756 &      2.818 &     -2.700 &        DBC \\
        65 &      25247 &      20.72 &      -0.33 &      4.130 &      2.067 &      2.063 &     -1.351 &        DDD &        140 &      72125 &      13.40 &      -0.10 &      4.604 &      2.334 &      2.270 &     -2.030 &        DCD \\
        66 &      25282 &      17.30 &      -0.28 &      5.066 &      2.975 &      2.091 &     -0.835 &        DCD &        141 &      72357 &       9.07 &      -0.24 &      5.225 &      2.985 &      2.240 &     -2.227 &        DCC \\
        67 &      26366 &      27.75 &      -0.63 &      4.094 &      1.806 &      2.288 &     -0.978 &        DCD &        142 &      72582 &      15.52 &      -0.29 &      5.472 &      2.876 &      2.596 &     -1.170 &        DCD \\
        68 &      26885 &      11.06 &      -0.55 &      4.897 &      2.212 &      2.685 &     -2.569 &        DDD &        143 &      72631 &      14.88 &      -0.39 &      4.934 &      2.797 &      2.137 &     -1.340 &        DDD \\
        69 &      27483 &      15.76 &      -0.27 &      4.509 &      2.247 &      2.262 &     -1.765 &        DDD &        144 &      73193 &      11.13 &      -0.07 &      5.509 &      3.122 &      2.387 &     -1.646 &        DCD \\
        70 &      27654 &      28.68 &      -0.75 &      3.756 &      1.405 &      2.351 &     -1.307 &        DDD &        145 &      73620 &      16.68 &      -0.22 &      4.390 &      2.111 &      2.279 &     -1.778 &        DCD \\
        71 &      27673 &      14.04 &      -0.14 &      3.975 &      1.522 &      2.453 &     -2.741 &        DCC &        146 &      73745 &      13.24 &      -0.35 &      4.516 &      1.672 &      2.844 &     -2.719 &        DCD \\
        72 &      28358 &      25.47 &      -0.15 &      3.719 &      1.380 &      2.339 &     -1.590 &        DCD &        147 &      73909 &      12.66 &      -0.49 &      5.240 &      2.878 &      2.362 &     -1.610 &        DCD \\
        73 &      28734 &      20.87 &      -0.01 &      4.156 &      2.179 &      1.977 &     -1.223 &        DCD &        148 &      74666 &      26.78 &      -0.44 &      3.461 &      1.223 &      2.238 &     -1.638 &        DCC \\
        74 &      29246 &       8.51 &      -0.54 &      5.352 &      2.898 &      2.454 &     -2.452 &        DCD &        149 &      75119 &      13.47 &      -0.14 &      5.352 &      2.619 &      2.733 &     -1.734 &        DDD \\
        75 &      29696 &      18.42 &      -0.33 &      4.319 &      1.712 &      2.607 &     -1.962 &        DCD &        150 &      75127 &      10.74 &       0.06 &      5.541 &      3.391 &      2.150 &     -1.454 &        DDD \\
\hline
\end{tabular} 
}
\end{table*}

\begin{table*}
\contcaption{Continued}
\setlength{\tabcolsep}{3pt}
\center
{\tiny
\begin{tabular}{ccccccccc|ccccccccc}
\hline
(1) & (2) &  (3) & (4) & (5) & (6) & (7) & (8) & (9) & (1) & (2) &  (3) & (4) & (5) & (6) & (7) & (8) & (9)\\
ID & Hip &    $\pi_0$ & $[M/H]$ & $V$ & $K_s$ & $V-K_s$ &$M_{K_s}$ & Quality & ID & Hip & $\pi_0$ & $[M/H]$ & $V$ & $K_s$ &$V-K_s$ & $M_{K_s}$ &    Quality \\
& &    (mas)  & (dex)  &(mag)& (mag) & (mag)   & (mag)    &         &    &     & (mas)   & (dex)   & (mag) & (mag) &(mag)   & (mag)     &  \\
\hline
       151 &      75458 &      32.23 &       0.03 &      3.290 &      0.671 &      2.619 &     -1.788 &        DBC &        218 &      97433 &      22.02 &      -0.47 &      3.841 &      1.732 &      2.109 &     -1.554 &        DCD \\
       152 &      76133 &      11.58 &      -0.13 &      5.497 &      2.929 &      2.568 &     -1.752 &        DCD &        219 &      97938 &      17.75 &      -0.32 &      4.715 &      2.171 &      2.544 &     -1.583 &        DCD \\
       153 &      76333 &      19.98 &      -0.42 &      3.914 &      1.524 &      2.390 &     -1.973 &        DDD &        220 &      98110 &      24.17 &      -0.09 &      3.886 &      1.371 &      2.515 &     -1.713 &        DCD \\
       154 &      76425 &      14.08 &      -0.17 &      5.264 &      2.975 &      2.289 &     -1.282 &        DCD &        221 &      98353 &      10.38 &      -0.38 &      4.837 &      2.718 &      2.119 &     -2.201 &        DCD \\
       155 &      76534 &      19.21 &      -0.55 &      5.245 &      3.103 &      2.142 &     -0.479 &        DCD &        222 &      98842 &      11.20 &      -0.63 &      4.991 &      2.034 &      2.957 &     -2.720 &        DDD \\
       156 &      76705 &      14.83 &      -0.34 &      4.663 &      2.290 &      2.373 &     -1.854 &        DDD &        223 &      98920 &      20.31 &      -0.03 &      5.094 &      2.708 &      2.386 &     -0.753 &        DCD \\
       157 &      77070 &      44.10 &       0.03 &      2.634 &      0.150 &      2.484 &     -1.628 &        CDD &        224 &      98962 &      18.97 &       0.12 &      5.399 &      2.723 &      2.676 &     -0.887 &        DDD \\
       158 &      77512 &      19.17 &      -0.32 &      4.594 &      2.668 &      1.926 &     -0.919 &        CCD &        225 &     100064 &      30.82 &      -0.18 &      3.576 &      1.466 &      2.110 &     -1.090 &        DDD \\
       159 &      77578 &      11.90 &      -0.28 &      5.207 &      2.929 &      2.278 &     -1.693 &        DDD &        226 &     100437 &       7.96 &       0.00 &      5.577 &      2.984 &      2.593 &     -2.511 &        CDD \\
       160 &      77853 &      19.36 &      -0.31 &      4.127 &      1.767 &      2.360 &     -1.798 &        DDD &        227 &     101101 &      16.29 &       0.03 &      4.913 &      2.504 &      2.409 &     -1.436 &        DDD \\
       161 &      78132 &      13.46 &       0.06 &      5.536 &      3.091 &      2.445 &     -1.264 &        DDD &        228 &     101936 &      13.87 &      -0.12 &      5.154 &      2.710 &      2.444 &     -1.580 &        CCD \\
       162 &      78159 &      14.72 &      -0.32 &      4.142 &      1.349 &      2.793 &     -2.811 &        DCC &        229 &     102014 &      13.74 &      -0.02 &      5.474 &      2.778 &      2.696 &     -1.532 &        DDD \\
       163 &      78481 &      12.76 &      -0.31 &      5.102 &      2.869 &      2.233 &     -1.602 &        DCD &        230 &     102453 &      16.21 &      -0.24 &      4.219 &      1.701 &      2.518 &     -2.250 &        DCC \\
       164 &      78650 &      15.69 &      -0.14 &      4.956 &      2.213 &      2.743 &     -1.809 &        DDD &        231 &     102488 &      44.86 &      -0.27 &      2.479 &     -0.007 &      2.486 &     -1.748 &        DCC \\
       165 &      79119 &      27.73 &      -0.20 &      4.730 &      2.338 &      2.392 &     -0.447 &        CCC &        232 &     102532 &      25.55 &       0.13 &      4.275 &      1.595 &      2.680 &     -1.368 &        DDD \\
       166 &      79302 &      10.21 &      -0.36 &      5.091 &      1.937 &      3.154 &     -3.018 &        CDC &        233 &     103004 &      17.20 &      -0.23 &      4.559 &      2.722 &      1.837 &     -1.100 &        DCD \\
       167 &      79882 &      30.63 &      -0.25 &      3.230 &      1.146 &      2.084 &     -1.423 &        DDD &        234 &     103519 &      12.58 &       0.00 &      5.553 &      3.529 &      2.024 &     -0.973 &        DDD \\
       168 &      80181 &      17.77 &      -0.08 &      4.857 &      2.615 &      2.242 &     -1.137 &        DCD &        235 &     103738 &      14.22 &      -0.23 &      4.670 &      2.541 &      2.129 &     -1.695 &        DCD \\
       169 &      80331 &      35.42 &      -0.21 &      2.732 &      0.579 &      2.153 &     -1.675 &        DBC &        236 &     104174 &       9.92 &      -0.01 &      5.204 &      2.747 &      2.457 &     -2.270 &        DCD \\
       170 &      80894 &      13.37 &       0.08 &      4.286 &      2.333 &      1.953 &     -2.036 &        DDD &        237 &     104459 &      20.46 &      -0.15 &      4.498 &      2.334 &      2.164 &     -1.111 &        DCD \\
       171 &      81437 &      11.98 &      -0.12 &      5.282 &      2.644 &      2.638 &     -1.964 &        CCC &        238 &     104732 &      22.77 &      -0.11 &      3.214 &      1.160 &      2.054 &     -2.053 &        CCC \\
       172 &      81833 &      30.02 &      -0.37 &      3.480 &      1.333 &      2.147 &     -1.280 &        DCD &        239 &     105411 &      12.03 &      -0.24 &      5.578 &      3.229 &      2.349 &     -1.370 &        DCD \\
       173 &      82396 &      51.19 &      -0.17 &      2.288 &     -0.392 &      2.680 &     -1.846 &        CCD &        240 &     105502 &      20.92 &      -0.14 &      4.081 &      1.752 &      2.329 &     -1.645 &        DCC \\
       174 &      82764 &      12.03 &      -0.26 &      5.386 &      3.128 &      2.258 &     -1.471 &        DDD &        241 &     105515 &      16.56 &      -0.23 &      4.279 &      2.206 &      2.073 &     -1.699 &        DDD \\
       175 &      83000 &      35.66 &      -0.03 &      3.188 &      0.728 &      2.460 &     -1.511 &        DCC &        242 &     106039 &      19.04 &      -0.28 &      4.498 &      2.263 &      2.235 &     -1.339 &        DDD \\
       176 &      84514 &      15.77 &      -0.03 &      4.716 &      2.040 &      2.676 &     -1.971 &        DCD &        243 &     106481 &      26.39 &      -0.31 &      3.982 &      1.901 &      2.081 &     -0.992 &        DDD \\
       177 &      85805 &      15.45 &      -0.20 &      5.073 &      2.426 &      2.647 &     -1.629 &        DDD &        244 &     106551 &      14.04 &      -0.08 &      4.869 &      2.595 &      2.274 &     -1.668 &        DCD \\
       178 &      86742 &      39.85 &       0.02 &      2.758 &      0.437 &      2.321 &     -1.561 &        DCC &        245 &     106944 &      14.37 &      -0.17 &      5.100 &      2.826 &      2.274 &     -1.387 &        DDD \\
       179 &      87194 &      15.37 &      -0.03 &      5.093 &      2.405 &      2.688 &     -1.662 &        CCD &        246 &     107119 &      17.84 &       0.04 &      4.552 &      1.722 &      2.830 &     -2.021 &        CBC \\
       180 &      87563 &       9.17 &      -0.12 &      5.174 &      2.021 &      3.153 &     -3.167 &        CCC &        247 &     107128 &      14.12 &      -0.01 &      5.240 &      2.901 &      2.339 &     -1.350 &        DDD \\
       181 &      87585 &      28.98 &      -0.09 &      3.731 &      1.045 &      2.686 &     -1.645 &        DCC &        248 &     107188 &      11.06 &      -0.20 &      4.720 &      2.549 &      2.171 &     -2.232 &        DCD \\
       182 &      87847 &       8.10 &      -0.16 &      5.441 &      2.946 &      2.495 &     -2.512 &        DDD &        249 &     108868 &      10.70 &      -0.11 &      5.550 &      3.521 &      2.029 &     -1.332 &        DDD \\
       183 &      87933 &      23.84 &      -0.10 &      3.702 &      1.491 &      2.211 &     -1.622 &        DCC &        250 &     109786 &      13.82 &      -0.13 &      5.330 &      3.364 &      1.966 &     -0.933 &        DDF \\
       184 &      88048 &      21.63 &       0.02 &      3.317 &      1.225 &      2.092 &     -2.100 &        DDD &        251 &     110003 &      17.39 &       0.01 &      4.167 &      2.061 &      2.106 &     -1.738 &        DDD \\
       185 &      88635 &      33.67 &      -0.36 &      2.984 &      0.545 &      2.439 &     -1.819 &        DDD &        252 &     110529 &      12.39 &      -0.07 &      5.526 &      3.143 &      2.383 &     -1.392 &        DCD \\
       186 &      88765 &      11.93 &      -0.10 &      4.645 &      2.426 &      2.219 &     -2.191 &        DCD &        253 &     110538 &      19.18 &      -0.39 &      4.418 &      1.880 &      2.538 &     -1.706 &        DCD \\
       187 &      88788 &       7.76 &      -0.06 &      4.996 &      2.778 &      2.218 &     -2.773 &        DDD &        254 &     110882 &      21.97 &      -0.37 &      4.777 &      2.420 &      2.357 &     -0.871 &        DDD \\
       188 &      89008 &       8.24 &      -0.14 &      5.571 &      3.648 &      1.923 &     -1.772 &        DDD &        255 &     111710 &      15.24 &       0.14 &      5.038 &      2.610 &      2.428 &     -1.475 &        DCD \\
       189 &      89065 &       8.02 &      -0.03 &      5.501 &      2.577 &      2.924 &     -2.902 &        DCD &        256 &     112067 &      11.72 &      -0.09 &      5.925 &      3.210 &      2.715 &     -1.445 &        CCD \\
       190 &      89153 &      12.71 &      -0.18 &      4.957 &      2.615 &      2.342 &     -1.864 &        DDD &        257 &     112529 &      12.80 &      -0.43 &      5.243 &      3.061 &      2.182 &     -1.403 &        DDD \\
       191 &      89772 &      21.35 &       0.05 &      5.406 &      2.904 &      2.502 &     -0.449 &        DCD &        258 &     112724 &      28.29 &      -0.12 &      3.497 &      1.276 &      2.221 &     -1.466 &        DCD \\
       192 &      89826 &      12.95 &      -0.09 &      4.335 &      1.838 &      2.497 &     -2.601 &        DCC &        259 &     112731 &       8.18 &      -0.53 &      5.428 &      2.625 &      2.803 &     -2.811 &        DCD \\
       193 &      89918 &      12.49 &      -0.21 &      4.846 &      2.791 &      2.055 &     -1.726 &        DDD &        260 &     112748 &      30.73 &      -0.16 &      3.512 &      1.181 &      2.331 &     -1.381 &        DCC \\
       194 &      89962 &      53.93 &      -0.42 &      3.234 &      1.050 &      2.184 &     -0.291 &        DDD &        261 &     113084 &      10.66 &      -0.02 &      5.818 &      3.514 &      2.304 &     -1.347 &        DDD \\
       195 &      90135 &      15.51 &      -0.17 &      4.663 &      2.545 &      2.118 &     -1.502 &        DDD &        262 &     113184 &      11.85 &       0.04 &      5.723 &      3.826 &      1.897 &     -0.805 &        DDD \\
       196 &      90139 &      27.40 &      -0.16 &      3.854 &      1.311 &      2.543 &     -1.500 &        DCC &        263 &     113246 &      21.14 &      -0.31 &      4.200 &      1.881 &      2.319 &     -1.493 &        DDD \\
       197 &      90344 &      10.34 &      -0.44 &      4.820 &      2.085 &      2.735 &     -2.842 &        DCC &        264 &     113521 &      10.89 &      -0.04 &      5.428 &      2.972 &      2.456 &     -1.843 &        DCD \\
       198 &      90496 &      41.72 &      -0.20 &      2.817 &      0.332 &      2.485 &     -1.566 &        DCD &        265 &     113864 &       7.87 &       0.07 &      5.252 &      2.325 &      2.927 &     -3.195 &        CCD \\
       199 &      90642 &      14.15 &      -0.08 &      5.381 &      2.944 &      2.437 &     -1.302 &        CCD &        266 &     113919 &      18.45 &      -0.20 &      4.640 &      2.154 &      2.486 &     -1.516 &        DCD \\
       200 &      91105 &       9.88 &      -0.07 &      5.116 &      3.022 &      2.094 &     -2.004 &        DCD &        267 &     114119 &      14.94 &      -0.14 &      4.483 &      2.351 &      2.132 &     -1.777 &        DDD \\
       201 &      92088 &      12.78 &      -0.12 &      4.833 &      2.222 &      2.611 &     -2.245 &        DCD &        268 &     114222 &      12.94 &       0.01 &      4.406 &      2.468 &      1.938 &     -1.972 &        DCD \\
       202 &      92689 &       6.79 &      -0.14 &      4.917 &      2.819 &      2.098 &     -3.022 &        DDD &        269 &     114641 &       9.40 &      -0.33 &      5.848 &      3.392 &      2.456 &     -1.742 &        DDD \\
       203 &      92782 &       9.39 &      -0.12 &      4.822 &      2.321 &      2.501 &     -2.816 &        DDD &        270 &     114855 &      21.75 &      -0.14 &      4.239 &      1.597 &      2.642 &     -1.716 &        DCD \\
       204 &      92872 &      11.02 &      -0.26 &      5.584 &      3.064 &      2.520 &     -1.725 &        DDD &        271 &     114971 &      23.64 &      -0.61 &      3.703 &      1.393 &      2.310 &     -1.740 &        DDD \\
       205 &      93026 &      16.10 &      -0.11 &      4.829 &      2.533 &      2.296 &     -1.433 &        DCD &        272 &     115088 &      16.03 &      -0.07 &      4.749 &      2.528 &      2.221 &     -1.447 &        DCD \\
       206 &      93244 &      20.96 &       0.00 &      4.024 &      1.786 &      2.238 &     -1.607 &        DCD &        273 &     115102 &      17.89 &      -0.22 &      4.405 &      1.718 &      2.687 &     -2.019 &        DCD \\
       207 &      93429 &      22.65 &      -0.19 &      4.017 &      1.636 &      2.381 &     -1.589 &        DDD &        274 &     115227 &       7.45 &      -0.63 &      5.052 &      1.993 &      3.059 &     -3.646 &        DCD \\
       208 &      93683 &      22.95 &      -0.11 &      3.755 &      1.573 &      2.182 &     -1.623 &        DDD &        275 &     115438 &      19.95 &      -0.40 &      3.960 &      1.468 &      2.492 &     -2.032 &        DDD \\
       209 &      93864 &      26.71 &      -0.17 &      3.324 &      0.458 &      2.866 &     -2.409 &        DDD &        276 &     115830 &      21.95 &      -0.12 &      4.273 &      1.859 &      2.414 &     -1.434 &        DCD \\
       210 &      94376 &      33.48 &      -0.27 &      3.072 &      0.883 &      2.189 &     -1.493 &        DCC &        277 &     115919 &      18.52 &      -0.03 &      4.537 &      2.442 &      2.095 &     -1.220 &        DDD \\
       211 &      94648 &      22.24 &       0.12 &      4.445 &      1.775 &      2.670 &     -1.489 &        DCD &        278 &     116076 &      12.61 &      -0.36 &      5.215 &      2.331 &      2.884 &     -2.165 &        CCD \\
       212 &      94779 &      26.27 &      -0.08 &      3.795 &      1.757 &      2.038 &     -1.146 &        DCC &        279 &     116591 &      12.16 &      -0.25 &      5.664 &      3.520 &      2.144 &     -1.055 &        DDC \\
       213 &      95498 &      11.98 &      -0.20 &      5.136 &      2.629 &      2.507 &     -1.979 &        DCD &        280 &     116853 &       9.21 &      -0.02 &      5.893 &      3.634 &      2.259 &     -1.545 &        DCD \\
       214 &      96229 &      30.30 &      -0.13 &      4.449 &      1.704 &      2.745 &     -0.889 &        DDD &        281 &     117073 &      14.45 &      -0.16 &      4.927 &      2.522 &      2.405 &     -1.679 &        CCC \\
       215 &      96683 &      12.23 &      -0.11 &      4.680 &      2.499 &      2.181 &     -2.064 &        DCD &        282 &     117314 &      10.63 &      -0.14 &      5.740 &      3.188 &      2.552 &     -1.679 &        DDD \\
       216 &      97118 &      11.27 &      -0.14 &      4.891 &      2.637 &      2.254 &     -2.103 &        DDD &        283 &     117375 &      10.27 &      -0.11 &      5.494 &      3.238 &      2.256 &     -1.704 &        DCD \\
       217 &      97290 &      15.70 &      -0.18 &      4.870 &      2.572 &      2.298 &     -1.449 &        DCD &        284 &     118209 &      13.89 &      -0.13 &      4.879 &      2.952 &      1.927 &     -1.334 &        DDD \\
\hline
\end{tabular} 
}
\end{table*}


\begin{thebibliography}{}

\bibitem[Alves(2000)]{Alves00}
Alves, D. R. 2000, ApJ, 539, 732 

\bibitem[Bahcall \& Soneira(1980)]{Bahcall80} 
Bahcall, J.N., Soneira, R.M. 1980, ApJS, 44, 73

\bibitem[Bilir et al.(2012)]{Bilir12a} 
Bilir, S., Karaali, S., Ak, S., {\"O}nal, {\"O}., Da{\v g}tekin, N.~D., Yontan, 
T., Gilmore, G., Seabroke, G.~M. 2012, MNRAS, 421, 3362 

\bibitem[Bilir et al.(2013)]{Bilir12b} 
Bilir, S., {\"O}nal, {\"O}, Karaali, S., Cabrera-Lavers, A., \c Cakmak, H. 2013, Ap\&SS, 344, 417

\bibitem[Cabrera-Lavers et al.(2005)]{Cabrera05}
Cabrera-Lavers, A., Garz{\'o}n, F., Hammersley, P. L. 2005, A\&A, 433, 173

\bibitem[Cabrera-Lavers et al.(2007a)]{Cabrera07a}
Cabrera-Lavers, A., Hammersley, P.L., Gonz\'alez-Fern\'andez, C., L\'opez-Corredoira, M.,  
Garz{\'o}n, F., Mahoney, T. J. 2007a, A\&A, 465, 825

\bibitem[Cabrera-Lavers et al.(2007b)]{Cabrera07b}
Cabrera-Lavers, A., Bilir, S., Ak, S., Yaz, E., L\'opez-Corredoira, M. 2007b, A\&A, 464, 565

\bibitem[Cabrera-Lavers et al.(2008)]{Cabrera08}
Cabrera-Lavers, A., Gonz\'alez-Fern\'andez, C.,Garz{\'o}n, F., Hammersley, 
P.~L., L\'opez-Corredoira, M. 2008, A\&A, 491, 781

\bibitem[Girardi et al. (2000)]{Girardi00} 
Girardi, L., Bressan, A., Bertelli, G., Chiosi, C. 2000, A\&AS, 141, 371 

\bibitem[Girardi \& Salaris(2001)]{Girardi01}
Girardi, L., Salaris, M. 2001, MNRAS, 323, 109

\bibitem[Grocholski \& Sarajedini(2002)]{Grocholski02}
Grocholski, A. J., Sarajedini A. 2002, AJ, 123, 1603 

\bibitem[Groenewegen(2008)]{Groenewegen08}
Groenewegen, M. A. T. 2008, A\&A, 488, 935

\bibitem[Karaali, Bilir \& Yaz G\"ok\c ce(2013)]{Karaali13}
Karaali, S. Bilir, S., Yaz G\"ok\c ce, E. 2013, PASA, 30, 11

\bibitem[Keenan \& Barnbaumet(1999)]{Keenan99}
Keenan, P.C., Barnbaum, C. 1999, ApJ, 518, 859 

\bibitem[Kubiak et al.(2002)]{Kubiak02}
Kubiak, M., McWilliam, A., Udalski, A., Gorski, K. 2002, AcA, 52, 159

\bibitem[Laney et al.(2012)]{Laney12}
Laney, C.~D., Joner, M.~D., Pietrzy{\'n}ski, G. 2012, MNRAS, 419, 1637 

\bibitem[Lopez-Corredoira et al.(2002)]{Lopez02}
L\'opez-Corredoira, M., Cabrera-Lavers, A., Garz{\'o}n, F., Hammersley, P. L. 2002, A\&A, 394, 883

\bibitem[Lopez-Corredoira et al.(2004)]{Lopez04}
L\'opez-Corredoira, M., Cabrera-Lavers, A., Gerhard, O., Garz{\'o}n, F. 2004, A\&A, 421, 953

\bibitem[Marshall et al.(2006)]{Marshall06}
Marshall, D.J., Robin, A.C., Reyl\'{e}, C., Schultheis, M., Picaud, S.
2006, A\&A, 453, 635

\bibitem[Paczynski \& Stanek(1998)]{Paczynski98}
P{\'a}czy{\'n}ski, B., Stanek, K. Z. 1998, ApJ, 494, L219 

\bibitem[Salaris \& Girardi(2002)]{Salaris02}
Salaris, M., Girardi, L. 2002, MNRAS, 337, 332

\bibitem[Sarajedini(1999)]{Sarajedini99}
Sarajedini, A. 1999, AJ, 118, 2321

\bibitem[Schlegel, Finkbeiner \& Davis(1998)]{Schlegel98}
Schlegel, D. J., Finkbeiner, D. P., Davis, M. 1998, ApJ 500, 525 

\bibitem[Skrutskie et al.(2006)]{Skrutskie06}
Skrutskie, M. F., et al. 2006, AJ, 131, 1163

\bibitem[Smith(1987)]{Smith87}
Smith, H. Jr., 1987, A\&A 171, 336

\bibitem[Stanek \& Garnavich (1998)]{Stanek98}
Stanek, K. Z., Garnavich, P. M. 1998, ApJ, 503, L131

\bibitem[Udalski(1998)]{Udalski98} 
Udalski, A. 1998, AcA, 48, 383 

\bibitem[van der Helshoecht \& Groenewegen (2007)]{vanHelshoecht07}
van der Helshoecht, V., Groenewegen, M.A.T. 2007, A\&A, 463, 559

\bibitem[van Leeuwen(2007)]{vanLeeuwen07}
van Leeuwen, F. 2007, A\&A, 474, 653

\bibitem[Zhao et al. (2001)]{Zhao01}
Zhao, G., Qiu, H.~M., Mao, S. 2001, ApJ, 551L, 85

\end{thebibliography}
\end{document}